# WheatAI v1.0: An AI-Powered High Throughput Wheat Phenotyping Platform

Maitiniyazi Maimaitijiang[1*], Hillson Ghimire[1], Subash Thapa[2], Mohammad Maruf Billah[1], Shaurya Sehgal[1], Mandeep Singh[2], Swas Kaushal[2], Kushal Poudel[1], Santosh Subedi[1], Ubaid Ur Rehman Janjua[1], Lise-Olga Makonga[1], Jyotirmoy Halder[2], Harsimardeep S. Gill [4], Mazhar Sher[3], Jagdeep Singh Sidhu[2], Sunish K. Sehgal[2*]

[1] Geospatial Sciences Center of Excellence, Department of Geography and Geospatial Sciences, South Dakota State University, Brookings, SD 57007, USA.
[2] Department of Agronomy, Horticulture, and Plant Science, South Dakota State University, Brookings, SD 57007, USA.
[3] Department of Agricultural and Biosystems Engineering, South Dakota State University, Brookings, SD 57007, USA.
[4] Department of Agronomy and Plant Genetics, University of Minnesota, St. Paul, MN 55108, USA

*  Corresponding authors: maitiniyazi.maimaitijiang@sdstate.edu
sunish.sehgal@sdstate.edu

**Abstract:**

High-throughput, low-cost phenotyping remains a critical bottleneck in wheat breeding, genetics, and crop management. This is particularly evident in the measurement of complex yield components (i.e., spike and spikelet counts), disease and grain-quality traits related to Fusarium Head Blight (FHB) and Fusarium-Damaged Kernels (FDK), and microscale physiological traits such as density and size of stomata and aperture. We introduce WheatAI (http://wheatai.net), an AI-powered web application designed to bridge the gap between advanced computer vision and AI/deep learning models, and high-throughput phenotyping (HTP) and practical agricultural applications. WheatAI v1.0 provides an accessible, browser-based interface that supports multiscale data ingestion from smartphones, Unmanned Aerial Vehicles (UAVs), and portable microscopes.

The platform's core functionalities include plot- and field-scale assessment via UAV- and smartphone-based wheat spike detection and counting, as well as smartphone-based spikelet counting. Additionally, it offers grain quality assessment through FDK ratio estimation and kernel morphometric measurements, such as length, width, and area, derived from smartphone images of kernel samples. For leaf-level analysis, WheatAI provides microscale phenotyping through automated stomatal counting, size, and aperture measurement from digital microscopy images. The system supports both single-image and bulk processing via a guided upload-and-run workflow. This platform is designed to reduce labor costs and rater subjectivity while accelerating field-to-lab decision cycles. By providing standardized, image-based outputs, WheatAI enables breeders, agronomists, and producers to implement high-throughput selection and precision scouting at scale.

## 1. Introduction

Wheat remains a cornerstone of global food security, yet meeting future demand under the dual pressures of population growth and increasing climatic volatility requires a significant acceleration in genetic gain [1]. While high-throughput genotyping has become a routine industry standard, phenotyping, which is the accurate and repeatable measurement of traits at scale, remains a persistent bottleneck [2]. Traditional breeding workflows still rely heavily on manual labor and subjective visual assessments [3, 4]. For instance, manually counting wheat spikes across large breeding nurseries and multilocation yield trials is labor-intensive and difficult to scale. Similarly, Fusarium Head Blight (FHB) and Fusarium-Damaged Kernels (FDK) scoring is often hindered by being highly time-consuming and subject to significant inter-evaluator variability [5, 6]. These limitations ultimately reduce the precision of genomic prediction and selection efficiency in breeding programs, while also constraining the accuracy and scale of field scouting and crop management practices [7, 8].

The convergence of low-cost, high-quality and accessible imaging devices/sensors and Artificial Intelligence (AI) provides revolutionary opportunities to enhance high-throughput phenotyping across multiple scales [9]. Modern imaging systems ranging from portable microscope and smartphones to low-cost Uncrewed Aerial Vehicles (UAVs) can capture high-resolution data of leaves, kernels, canopies, and entire fields rapidly [10]. This allows for the automated measurement of critical wheat traits, including macro-level yield components such as spikes per unit area (SN) and spikelets per spike (SPS), as well as microscale physiological traits such as stomatal density (SD) and aperture [11, 12].

However, a significant gap persists due to the lack of low-cost, robust, and operational phenotyping models and pipelines. While many AI model applications are published as proof-of-concepts, they often lack of scalability and transferability and fail to translate into accessible platforms and tools that can handle the throughput of real-world breeding programs, commercial scouting, and stakeholder usage [13, 14]. Without robust and scalable AI models, alongside integrated lab- and field-ready platforms, breeders, geneticists, pathologists, physiologist, agronomists, and producers remain without the necessary infrastructure to implement AI-driven selection and precision management at scale [4, 15, 16].

To address this gap, we developed WheatAI v1.0 (http://wheatai.net), an AI-powered web platform that packages multiple wheat phenotyping and disease or grain-quality AI models developed by our team, into a unified, accessible workflow. WheatAI is designed not only to provide model inference but also to support end-to-end phenotyping operations including data ingestion, trait extraction, visualization, and export into a unified workflow. By transitioning complex AI pipelines into a user-friendly browser interface, the platform empowers researchers, agronomists, and producers to transform raw imagery into standardized and actionable biological insights for both selection and in-field decision-making.

## 2. WheatAI Platform Overview and Functional Capabilities

The design of WheatAI followed key principles such as scalability, accessibility, and interoperability. The platform minimizes end-user computing requirements by adopting a cloud-native user-server architecture. The front end of the platform (***Figure 1***) is a lightweight, browser-based interface built using open-source HTML, CSS, and JavaScript frameworks to ensure cross-platform access from office workstations, laptops, and field devices. The frontend is implemented as a decoupled single-page application (SPA) built with a Vite-based JavaScript framework, enabling fast user-side rendering and asynchronous communication with the backend via RESTful APIs. The backend utilizes a Python and Django REST API to provide secure endpoints connecting to GPU-accelerated PyTorch inference modules supporting both rapid single-image inference and a distributed task queue for asynchronous high-volume batch processing. It performs data management and AI inference using scalable compute resources, such as institutional servers or cloud services, allowing the system to process large image collections and high-resolution products without degrading the user experience. Inference tasks are executed asynchronously using Celery workers connected through Redis, allowing multiple detection jobs to be processed in parallel without blocking user interactions and handling long-running inference tasks efficiently. Deployment is containerized using Docker, with Nginx, Gunicorn, and environment-specific configuration ensuring a production-ready stack for wheatai.net.

The images can be collected from using open source apps like Field Book [17] or KDSMart for structure naming and linkage to the genotype ID or plot ID. Users can upload collected imagery, adjust detection parameters, and review results through an intuitive browser-based interface built with modern JavaScript frameworks. WheatAI v1.0 is designed for both rapid and scalable use cases by supporting single-image processing for quick checks and multi-image bulk processing for high-throughput, trial-scale analysis. Outputs are exportable as processed or annotated images and summarized statistics in open-standard Excel or CSV formats allowing interoperability for direct uploading into BrAPI enabled databases like T3 (https://wheat.triticaetoolbox.org/) and enable downstream breeding records and data-driven decision-making.

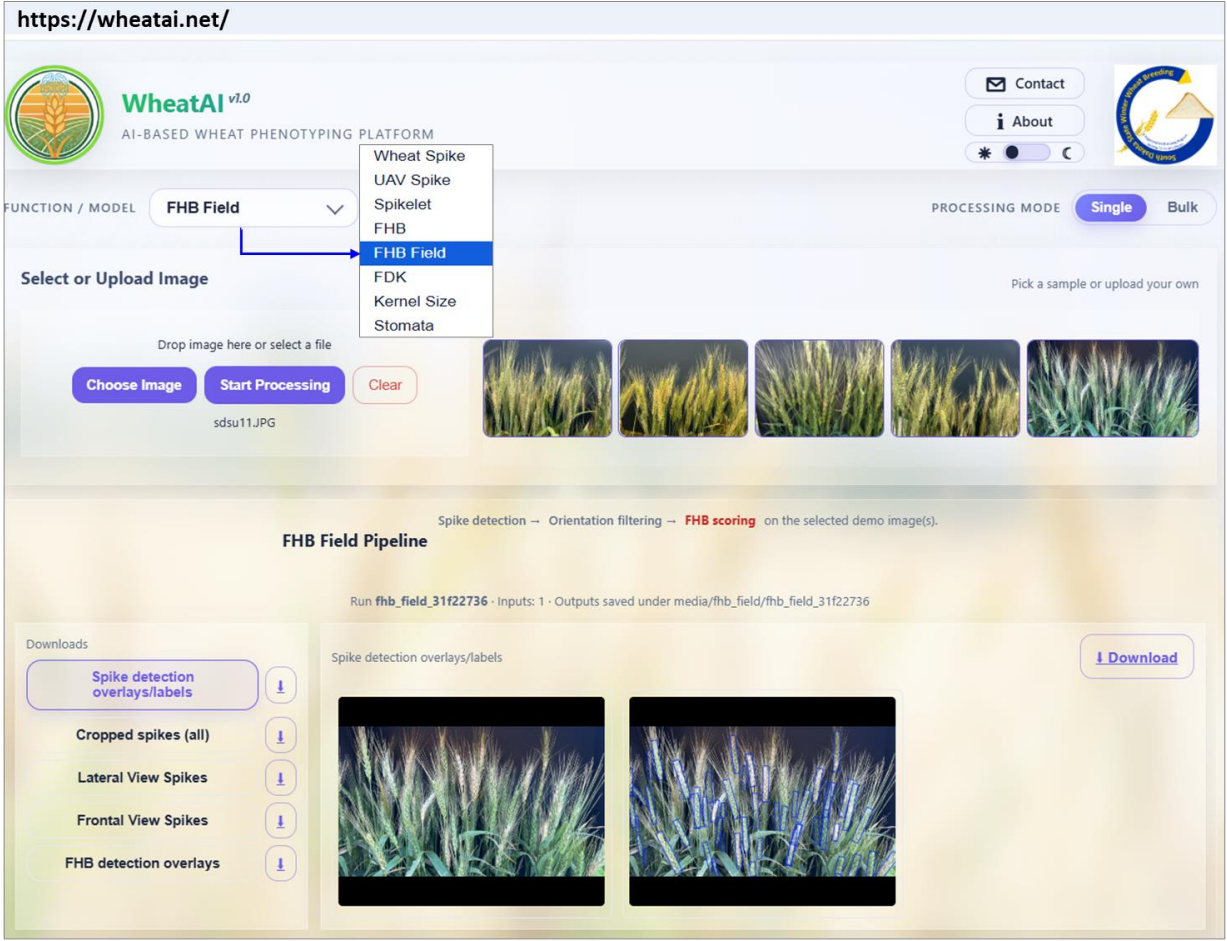

**Figure 1.** WheatAI v1.0 web interface and FHB-Field workflow example. Screenshot of the WheatAI (wheatai.net) browser-based platform showing the model-selection dropdown (Wheat Spike, Spikelet, FHB, FDK, Kernel, and Stomata), single/bulk processing options, and the guided image upload/demo selection. The example illustrates the FHB Field pipeline results, including spike detection overlays/labels, cropped spike outputs (lateral/frontal views), and FHB detection overlays, all available for download for downstream analysis and reporting.

## 2.1 Module 1: Wheat Spike Detection and Counting

The number of spikes per unit area serves as a critical wheat yield component and a primary parameter for the assessment of yield potential [18]. WheatAI facilitates automated spike detection and counting for both close-range smartphone or tablet RGB imagery, including top-view and side-view perspectives, as well as low-altitude UAV RGB imagery. The platform architecture supports object-detection-based inference with visual overlays and exportable count metrics, which enables the rapid extraction of yield component traits from standard RGB images.

To ensure high precision across varying field conditions and plant densities, this module employs advanced object detection architectures, specifically variants of the YOLO family such as YOLOv12 [19]. By utilizing oriented bounding boxes (OBB) [20] rather than standard horizontal boxes, the model can more accurately delineate spikes that are tilted or overlapping, which is a common occurrence in dense wheat canopies. This approach significantly reduces counting errors and provides a more robust foundation for yield forecasting and high-throughput selection.

## 2.2 Module 2: Wheat Spikelet Detection and Counting

The quantification of spikelets per spike is a fundamental determinant of grain sink strength and a vital indicator of a genotype's reproductive potential [21]. This module is designed to isolate and count these individual floral structures from high-resolution imagery captured in either controlled laboratory settings or via side-view field-canopy photography using smart phone or tablet. By automating the identification of these dense subunits, WheatAI facilitates a deeper investigation into the yield components that contribute to overall grain number without the bottlenecks associated with manual dissection. The technical framework utilizes YOLOv12 with oriented bounding boxes (OBB) to navigate the complex, angular arrangement of spikelets along the rachis. This specific OBB implementation allows the model to precisely align with the natural tilt of each spikelet, effectively distinguishing between overlapping structures and reducing errors caused by the irregular orientation of the wheat head within the canopy.

## 2.3 Module 3: Wheat Fusarium Head Blight (FHB) Assessment

Fusarium Head Blight (FHB), primarily caused by *Fusarium graminearum*, represents a major threat to global wheat production due to its impact on grain yield and the accumulation of mycotoxins such as deoxynivalenol [22]. Rapid, accurate, and low-cost assessment of wheat FHB at scale is critical for accelerating the selection of resistant breeding lines [23], and enabling timely, field-based assessments that support production decisions and help mitigate downstream losses [24]. WheatAI addresses the need for rapid FHB assessment through two specialized submodules tailored to different phenotyping scales. For controlled environments, a single-spike submodule utilizes a unified YOLOv12 architecture for the simultaneous detection and classification of disease symptoms on individual heads. For field-based applications, a more complex canopy submodule processes side-view imagery through a multi-stage pipeline. This field pipeline first performs spike detection and extraction, followed by a lightweight 3D CNN adapted for single-image binary classification for spike orientation and quality selection, and finally employs YOLOv12 with oriented bounding boxes (OBB) to identify and classify symptomatic spikelets. Together, these components enable standardized, non-destructive FHB quantification across laboratory and field conditions.

By integrating these advanced architectures, the platform can accurately calculate disease incidence, severity, and overall disease index from both lab and field images. The use of OBB in the field submodule is particularly critical as it allows the system to isolate specific infected regions on a spike regardless of its angle in the canopy, ensuring standardized disease metrics are extracted even under challenging environmental conditions. This dual-layered approach provides a rapid, non-destructive assessment of FHB resistance that is suitable for both high-throughput breeding trials and real-time field monitoring.

## 2.4 Module 4: Wheat Fusarium Damaged Kernels (FDK) Assessment

Wheat Fusarium-Damaged Kernels (FDK) assessment is a critical post-harvest process used to evaluate grain quality and safety, as infected kernels often harbor dangerous levels of mycotoxins [25]. WheatAI streamlines this evaluation by supporting FDK assessment from smartphone images of wheat kernel samples, effectively replacing the labor-intensive and error-prone process of manual sorting. The module utilizes the YOLOv12 architecture with oriented OBB to perform simultaneous kernel detection, counting, and health-status categorization. By applying OBB, the system can precisely isolate each kernel even when they are densely packed or randomly oriented on a sampling surface, allowing for the accurate and rapid calculation of FDK ratios. This automated post-harvest screening provides breeders and grain handlers with a standardized, objective metric for grain-quality assessment, significantly reducing the time required for sample processing.

## 2.5 Module 5: Wheat Kernel Morphometrics

The precise measurement of grain physical characteristics, such as length, width, and surface area, is essential for understanding wheat yield and end-use quality [26]. WheatAI automates this morphometric analysis by processing images of grain samples and extracting detailed geometric traits at scale. To transform image-based pixel dimensions into precise physical measurements in millimeters, the system utilizes ArUco markers as high-contrast fiducial references included within the imaging field. These markers provide a standardized scale calibration, which is essential for maintaining accuracy across different camera heights and lens focal lengths.

The computational module utilizes a two-stage computer vision pipeline that begins with the YOLOv12 architecture for initial kernel detection with oriented bounding boxes (OBB). Once the kernels are localized, the system integrates the pre-trained zero-shot Segment Anything Model (SAM) [27] to perform high-fidelity instance segmentation within those boxes. By applying the pre-trained SAM guided by YOLO-detected bounding box prompts, the model can precisely delineate the boundary of each kernel for kernel area calculation and account for irregular shapes or surface textures that traditional horizontal bounding boxes often miss. This combination of ArUco-based scale calibration, YOLOv12 detection, and SAM-based segmentation allows for the generation of standardized morphometric data that is highly repeatable, providing breeders and researchers with a robust tool for studying the genetic basis of grain size and shape.

## 2.6 Module 6: Wheat Stomata Morphometrics

The automated extraction of microscale leaf traits, such as stomatal and pore aperture size and density, is critical for understanding the physiological responses of wheat to environmental stressors like heat and drought [28]. WheatAI includes a dedicated module that supports standardized phenotyping of stomatal and pore aperture size and density from handheld, portable digital microscope images acquired at 400× magnification. This module utilizes the integrated scale bars provided by the microscope imaging system to establish the necessary pixel-to-millimeter or micrometer ratio. The computational workflow employs a two-stage pipeline beginning with the YOLOv12 architecture to detect both stomata and their respective pores within OBB. Once localized, the pre-trained Segment Anything Model (SAM) is applied within these boxes to precisely delineate boundaries for area and aperture measurements.

By combining the high-speed detection capabilities of YOLOv12 with the high-fidelity segmentation of SAM, the platform can accurately distinguish between open and closed stomata even in images with complex textures or varying contrast. This integration allows for the rapid generation of standardized morphometric data, providing researchers with a reliable tool for investigating leaf gas exchange and water-use efficiency at scale.

All outputs from the above-mentioned six functional modules are exportable as high-resolution annotated images and comprehensive summarized statistics in Excel or CSV formats to support breeding records and data-driven agronomic decisions. The platform also provides interactive visualizations within the browser interface, allowing users to verify model performance and audit detected features before finalizing their datasets. By streamlining the transition from raw imagery to structured biological information, WheatAI effectively bridges the technical gap between advanced deep learning and operational agricultural practice.

## 3. Limitations of Current Version and Future Work

While WheatAI v1.0 represents a significant step toward operationalizing AI for wheat phenomics, several technical challenges define our roadmap for future development. A primary concern is that AI model transferability remains relatively limited because training datasets often lack the extensive variations in sample size and environmental diversity required for universal application. This domain shift is particularly evident in field-based FHB assessment and kernel-based FDK evaluation, where model performance can fluctuate across different cultivars, growth stages, camera specifications, and illumination conditions. Furthermore, occlusion and clutter within dense canopies continue to be a significant hurdle, as overlapping spikes and spikelets can impede the accuracy of detection and segmentation methods. Beyond these biological and environmental factors, inference speed and computational costs may vary depending on image resolution, batch size, and available server resources, which can impact the feasibility of real-time deployment in bandwidth-limited field environments. These factors, combined with acquisition variability such as motion blur and inconsistent sampling geometry, can degrade the stability of automated trait extraction.

Future iterations of the platform will focus on enhancing model robustness and transferability through training on larger, high-quality datasets that encompass a wider range of environmental conditions and wheat varieties. To improve accessibility for field-based operations, a mobile application framework utilizing Mobile Cloud Computing (MCC) will be implemented to allow for real-time inference directly from the field. From a systemic perspective, upcoming versions will incorporate essential enterprise-level features including user authentication, secure registration, and robust database management to allow users to store and track their longitudinal phenotyping data. Data governance will also be a priority, with the implementation of clear communication regarding data privacy and storage policies. Finally, we intend to integrate a discussion panel within the platform to foster collaboration and troubleshooting among the community of breeders, geneticists, pathologists, physiologists, agronomists, and other stakeholders.

## 4. Conclusion

WheatAI v1.0 presents a unified, cloud-native web platform that operationalizes AI-driven, high-throughput wheat phenotyping across multiple spatial scales, ranging from field-level yield components to microscale physiological traits. By integrating six specialized modules-including spike and spikelet phenotyping, Fusarium Head Blight (FHB) and Fusarium-Damaged Kernel (FDK) assessment, kernel morphometrics, and stomatal analysis WheatAI transforms raw imagery from smartphones, UAVs, and portable microscopes into standardized, actionable biological information through an accessible end-to-end workflow.

A key contribution of this work lies in bridging the persistent gap between advanced AI model development and real-world agricultural deployment. Rather than introducing isolated proof-of-concept algorithms, WheatAI demonstrates how robust deep learning methods can be embedded within an integrated, browser-based platform capable of supporting both single-image inference and high-throughput, bulk processing required by modern wheat breeding programs and operational scouting. This design substantially lowers technical barriers for breeders, geneticists, pathologists, physiologist, agronomists, and producers, enabling broader adoption of AI-driven selection and precision management practices.

While the current release represents an important step toward scalable phenotyping infrastructure, ongoing development will focus on improving model robustness across diverse environments and cultivars, expanding training datasets, and enhancing system-level features such as mobile cloud computing, secure user authentication, and long-term data management. Together, these advancements will further strengthen WheatAI's role as a practical, extensible platform for accelerating genetic gain, improving disease management, and supporting data-driven decision-making in wheat production systems.

**ACKNOWLEDGMENTS**

This work was supported in part by the South Dakota Wheat Commission, the U.S. Wheat and Barley Scab Initiative (USWBSI, No. 59-0206-2-153), the USDA–NIFA Wheat Coordinated Agricultural Project (Award No. 2022-68013-36439), and the South Dakota Nutrient Research and Education Council (NREC). Additional support was provided by the USDA–NIFA Hatch Project (No. SD00H757 and SD00H829-25) through the South Dakota Agricultural Experiment Station at South Dakota State University. This project was also supported by the National Science Foundation under Award No. 2316502.